\documentstyle[preprint,aps,floats,epsf]{revtex}
\firstfigfalse
\def\mkfig#1{}

\def\mkfig#1{#1}

\def\intl{\int\limits}
\def\lab#1{\label{#1} }

\begin{document}

\draft

\preprint{ENSLAPP A-488/94, hep-ph/9410247}

\title{On the consistent solution of the gap--equation \\ for
spontaneously broken $\lambda \Phi^4$-theory  }

\author{Herbert Nachbagauer\footnote[3]{e-mail: herby @ lapphp1.in2p3.fr} }

\address{  Laboratoire de Physique Th\'eorique ENSLAPP
% {\small E}N{\large S}{\Large L}{\large A}P{\small P}
\footnote{URA 14-36 du CNRS, associ\'ee \`a l'E.N.S. de Lyon,
et au L.A.P.P. (IN2P3-CNRS)\\
\hspace*{0.7cm} d'Annecy-le-Vieux} \\
Chemin de Bellevue, BP 110, F - 74941 Annecy-le-Vieux Cedex,
France}

\date{\today}

\maketitle

\begin{abstract}
We present a self--consistent solution of the
finite temperature
gap--equation for $\lambda \Phi^4$ theory beyond the Hartree-Fock
approximation
using a composite operator effective action. We find that in a
spontaneously broken theory not only the so--called daisy and
superdaisy graphs contribute to the resummed mass, but
also resummed non--local diagrams are of the same order,
thus altering the effective mass for small
values of the latter.
\end{abstract}

\pacs{} % no pacs numbers }

\newpage

\section*{Introduction }

The study of $\Phi^4$ -- theory at finite temperature
is of great interest for a wide field
of applications. A self--interacting scalar field
serving as a simple model for the Higgs particle in the standard  model of
electroweak interactions may allow the study of  symmetry changing phase
transitions. In fact, the order of the electroweak phase transition
plays a crucial r\^ole in
the framework of cosmological scenarios as well as for
the badly understood process of baryogenesis \cite{kolbt}.
Despite the simplicity of the model, one may at least hope to gain
some insight in the mechanism  of the phase transition. Moreover, the
theory is a suitable test ground for analytic non--perturbative
methods, e.g. variational methods \cite{stev} as well as for lattice
simulations \cite{latt}.

High temperature symmetry restoration in a
spontaneously broken theory was already noted by Kirzhnits and Linde
\cite{kirzl}
and worked out quantitatively subsequently \cite{dwk}.
The convenient tool to study the behavior of the theory turns out
to be the effective potential. Whereas the
critical temperature is
already determined by the one loop potential and thus relatively
simple to find, the order of the phase transition depends
on the detailed shape of the potential which requires also
the analysis of
higher loop contributions, even for small coupling constant.

In particular, one finds, quantizing the theory around the classical
non--trivial minimum that the one loop self--energy at high temperature
behaves like $(m_T)^2 \sim (\sqrt{\lambda} T)^2 $, independent of external
momenta and the value of the classical minimum. Thus, for large $T$
the thermal
mass dominates over the tree-level mass and the minimum of the
effective action becomes the trivial one.
The theory exhibits two important features: non--temperature stable vacuum
and effective temperature dependent mass. It was consequently proposed
\cite{stev}  to move
the tree level mass into the interaction part  and to
start perturbation theory with the free Lagrangian
%$\onehalf %
${1\over 2}
\Phi (\Box  + \Omega^2)  \Phi $ including a yet undetermined mass
parameter $\Omega$.  Then one
calculates the effective  potential and fixes the parameter by the 'principle
of minimal sensitivity' $\partial V (\Phi_{min}) / \partial \Omega =0 $.

A more systematic approach of a self--consistent loop expansion was developed
already some time ago by Cornwall, Jackiw, and Tomboulis (CJT) in their
effective action formalism for composite operators \cite{CJT}. The basic idea
is to  introduce a bilocal mass operator in the generating
functional instead of the local mass $\Omega$. Then one defines
the generalized effective action $\Gamma$ as the double Legendre transform of
the generating functional which is now not only a
functional of the expectation value of the field but also depends on the
expectation value of the time--ordered product $T\Phi(x) \Phi(y)$.
The principle of minimal sensitivity gets replaced by the
the so--called gap--equation
$\delta \Gamma (\Phi,G) / \delta G(x,y) = 0 $ which is employed
to eliminate the exact two--point Green function $G(x,y)$.

This paper is dedicated to investigate on the solution of the gap--equation
beyond the Hartree-Fock approximation. Whereas the latter amounts to simply
include only a local mass term in the ansatz for the  Green function
\cite{pi,8},  thus being a slightly
more elegant reformulation of the variational  approaches mentioned
above, we will consistently \cite{9}
include also the non-local contribution appearing in
a spontaneously broken theory.

Our approach is as follows.
We split the inverse Green function into a
local mass term absorbing the usually resummed graphs,
and a  non-local self--energy part. By this ansatz,
the gap--equation is recast into a non--linear integral equation for
the self--energy, which we solve approximately  by
a suitably chosen ansatz.
Self consistency requires the unknown mass parameter to satisfy an
'effective-mass'  equation, the latter
differing  from the one found for  the effective mass in the superdaisy
resummation by an additional  term  which dominates in the limit of
vanishing mass parameter.   We discuss the behavior of the mass parameter
near the critical temperature.   We lay great emphasis on dealing with
the divergencies correctly.

\section*{CJT composite operator formalism }

It is useful to briefly review the basic steps in the construction of the
generalized
effective action for a classical action $I(\Phi) $ \cite{CJT}.
One starts with the generating  functional in Euclidean  space--time
$$ Z(J,K) =\int \cal D \Phi \exp  -
\left[
I(\Phi) + \intl_x J(x) \Phi(x) +
{1 \over 2} \intl_x\intl_y \Phi (x) K(x,y) \Phi(y) )\right] $$
and $ W(J,K) = \log Z(J,K) $.
After defining the classical field $\phi (x)$ and the Green function
$G(x,y)$ by
\begin{equation}
- {\delta W (J,K) \over \delta J(x) }  = \phi (x) = \left< \Phi(x)
\right>_{J,K}
, \qquad
- { \delta W (J,K) \over \delta K(x,y) }= {1\over 2} (G(x,y) + \phi (x) \phi(y)
)
\lab{2}
\end{equation}
one performs a Legendre transformation to find the generalized effective
action
$$ \Gamma (\phi ,G ) = W(J,K) + \intl_x \phi (x) J(x) + {1 \over 2}
\intl_x \intl_y \left( \phi (x) K(x,y) \phi (y) + G(x,y) K(x,y) \right)
$$
where $J,K$ at the r.h.s.\ have to be taken at the solution
of Eq. (\ref{2}). From the above definition, one immediately deduces the
inverse relations
$$ { \delta \Gamma (\phi ,G) \over  \delta
\phi(x) } = J(x) + \intl_y K(x,y) \phi(y) ,
\qquad
 { \delta \Gamma (\phi,G) \over \delta G(x,y) }  = { 1 \over 2 } K(x,y ) .$$
The last equation tells us that if one wants back the standard
effective action defined by $K(x,y) = 0 $ one has to impose the
so--called 'gap--equation'
$$ { \delta \Gamma ( \phi,G) \over \delta G(x,y) } = 0 .$$
In order to give a series expansion of $\Gamma(\phi,G)$ we define the
functional operator
\def\tr{\mbox{Tr}}
$$ D^{-1 }(\phi;x,y) = { \delta^2 I(\phi )\over \delta \phi(x)
\delta \phi(y) } $$
and one finds
$$ \Gamma (\phi,G) = I(\phi) +{1 \over 2 } \tr \ln ( D_0 G^{-1} )
+{1 \over 2 } \tr ( D^{-1} G - 1 ) + \Gamma^{(2)} (\phi,G) ,$$
$ D_0 (x,y )$ being the free propagator derived from the part of
the action which is quadratic in the fields.
The quantity $ \Gamma^{(2)} (\phi,G) $ contains all two--loop
contributions and higher and has to be calculated as follows:
Shift the field $\Phi$ in the classical action by $\phi$.
Then $I(\Phi + \phi )$ contains terms cubic and higher in $\Phi$
which define the vertices. $\Gamma^{(2)} (\phi,G) $ is
given by all two-particle-irreducible vacuum graphs with
the given set of vertices and the propagators replaced by
$G(x,y)$. The gap--equation  reads
$$ G^{-1} (x,y) = D^{-1} (x,y) + 2 {\delta \Gamma^{(2)} (\phi,
G) \over \delta G(x,y ) } .$$

\section*{ $\lambda \Phi^4 $--theory }

\mkfig{
\begin{figure}
\epsfxsize4in
\hspace*{\fill}
\epsfbox{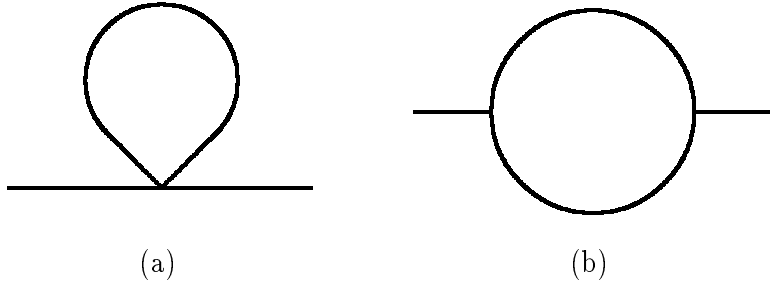}
\hspace*{\fill}
\vspace{0.3cm}
\caption{Feynman graphs contributing to the lowest order self--energy.
Lines denote the full Green function $G(x,y)$, the tree--vertex is
given by $\lambda\phi/3!$ and the four--vertex by $\lambda / 4!$\, .}
\end{figure}
\vspace{0.5cm}
}

The self--energy graphs we are going to resum are shown in Fig.~1.
Although the non--local contribution  Fig.~1b is
formally of order $\lambda^2$ it contributes to order $\lambda$
in the false vacuum since already the tree--level value of the
expectation value of the field is
$ \phi \sim  m/\sqrt{\lambda} $. Consequently, the three--vertex
$\lambda \phi /3! $ is
of order $\sqrt{\lambda} $ which means that
in the case of a spontaneously broken theory
the non--local contribution (b) is as important as the local one (a).
We will show that this is also true at finite temperature.
Thus, the Hartree--Fock approximation involving only the local
graph which corresponds  to the superdaisy  resummation
at finite temperature is incomplete.
Including the non--local graph Fig.~1b  in the
self--energy, the gap--equation in Fourier--space reads

\goodbreak

\begin{eqnarray}
 G^{-1} (\vec k,\omega)=\vec k^2 + \omega^2 +
\overbrace{ (-m^2 + {\lambda \over 2}
\phi^2 ) + {\lambda \over 2 } T \sum_{\omega '} \int {d^3 p \over
(2 \pi)^3 } G(\omega ',\vec p ) }^{M^2}  \; +
& & \nonumber \\
 + { 1 \over 2 } ( \lambda \phi)^2  T
\sum_{\omega '} \int {d^3 p \over (2 \pi)^3 }
G(\omega,\vec p) G(\omega - \omega ',\vec p - \vec k )
\lab{5}
\end{eqnarray}
where we use imaginary time formalism with
$ \omega = 2 \pi n T ,\; n \in {\bf Z } $.
The first sum/integral corresponding to graph (a) does not depend on
external momenta and energies and thus can be treated as  mass term.
We employ  the ansatz
\begin{equation}  G^{-1} (\omega,\vec k) = \vec k^2 + \omega^2 + \mu^2 +
\Pi(\omega,\vec k)  \lab{ans1}
\end{equation}
and subsequently absorb all local contributions in the mass parameter
$\mu$ which has to be determined self--consistently afterwards.
Plugging (\ref{ans1}) into  (\ref{5}), the gap--equation transforms
into an equation for $\Pi(\omega,\vec k)$,
\begin{eqnarray}
\Pi(\omega,\vec k) =
M^2 - \mu^2 +  {1 \over 2 } ( \lambda \phi)^2 T
\sum_{\omega '} \int {d^3 p \over (2 \pi)^3 }
{1 \over \vec p^{\, 2} + {\omega '}^2 + \mu^2 + \Pi(\omega ',\vec p) } \times
& & \nonumber  \\
\times{ 1 \over (\vec k - \vec p\, )^2 + \mu^2 + (\omega - \omega ')^2 + \Pi
(\omega - \omega ' , \vec  k - \vec p) }
. \lab{6} \end{eqnarray}
Simple power counting now reveals that the expression involving the integration
is logarithmic divergent for large momenta $\vec p$.
To isolate these divergencies,
we subtract and add a term of the form
\begin{equation}
T \sum_{\omega '} \int {d^3 p \over (2 \pi)^3 }
{ 1 \over \vec p^{\, 2} + {\omega '}^2 } { 1 \over \vec p^{\, 2} +
{\omega '}^2 +  \mu^2 }
\lab{subtr}
\end{equation}
which in the high temperature $(T \gg \mu)$ limit evaluates to
\begin{equation}
{T \over 4 \pi \mu } -  {\cal D} ,
  \quad {\cal D} =
{1\over 8 \pi^2 }( 1 - \gamma + \log  {2 \pi} + \log{
 T \over \Lambda }  ) , \lab{diva}
\end{equation}
where we introduced a UV--cutoff $\Lambda$.
The difference between the last summand in Eq. (\ref{6}) and the subtraction
(\ref{subtr}) is a UV--finite quantity. Consequently, in the high
temperature limit, we are allowed to pick only the zero--mode in the
thermal sum in the difference (\ref{6}) - (\ref{subtr}).
The $\omega' =0$ contribution of the subtraction exactly
cancels the first term in (\ref{diva}), leaving us with
${1\over 2} \lambda^2 \phi^2 \cal D$ as
divergent part of the r.h.s.\ of the gap--equation (\ref{6}).
It is now possible to absorb the logarithmic divergence in the mass
renormalization and we choose the parameter $\mu$ such that
\begin{equation}
M^2 - \mu^2 - {\lambda^2 \phi^2 \over 2 } {\cal D } = 0
\lab{2a} \end{equation}
which renders the self--energy $\Pi ( \omega ,\vec k) $ in equation
(\ref{6}) a finite quantity.
Furthermore, one may  simplify equation (\ref{6})
by considering the  $\omega  =0$ mode of the self--energy only
since the non--static modes are at least
suppressed by a factor $1 / T^2 $.   Thus putting
$\Pi( \omega=0, | \vec k| ) = \Pi(k)$ we end  with the integral equation
\begin{equation}
\hat \Pi(\hat  k ) = \kappa  \intl_0^\infty d\hat p \hat p^2
\intl_{-1}^1 {dz \over ( \hat p^2 + 1 + \hat \Pi( \hat p ) ) ( \hat p^2 +
\hat k^2
+ 2 \hat p \hat k z + 1  + \hat \Pi (\sqrt{\hat p^2 + \hat k^2 + 2 \hat p
\hat k z } ) ) }
\lab{7a}
\end{equation}
where we introduced the dimensionless quantities
$p =    \mu \hat p, \;  \Pi (p) =  \mu^2 \hat \Pi (\hat p) $ and the parameter
$\kappa =  T (\lambda \phi )^2 / ( 8 \pi^2 \mu^3 ). $

\section*{Solution of the gap--equation}

Without  going in the details of solving (\ref{7a}) one can already
make some useful statements about the behavior of $\hat\Pi (\hat k)$.
Firstly, the denominator of (\ref{7a}) is a positive definite quantity,
thus $\hat \Pi (\hat k) \ge 0 $.
Furthermore, assuming that the solution is bound,
i.e. $\hat \Pi ( \hat k) \le \hat \Pi_{max}$, the denominator may be
 estimated form
below (above) by $(\hat p^2 + 1)(\hat p^2+\hat k^2+2\hat p\hat k z +1)$ and
$(\hat p^2+1+\hat\Pi_{max} )(\hat p^2+\hat k^2+2 \hat  p \hat k z + 1 + \hat
\Pi_{max} )$ respectively.
Carrying out the integration on the r.h.s.\ in (\ref{7a}),
one finds the following  bounds for $\hat \Pi (\hat k)$
\begin{equation}
 {\pi \over \hat k} \arctan {\hat k \over 2 \sqrt{1 + \hat \Pi_{max} } }
\le {\hat \Pi ( \hat k) \over \kappa }  \le {\pi \over \hat k} \arctan
 {\hat k \over 2 }
 \lab{bounds}
\end{equation}
which in turn determine the asymptotic behavior,
$\hat \Pi ( \hat k ) =  \pi^2 \kappa / (2 \hat k )  + {\cal O}( \hat k^{-2} )
 $.

Secondly, we note that due to the z--integration, odd powers of z
vanish in the integrand, which in particular means that $ \hat \Pi'(0) =0.$

Using this property and the asymptotic behavior, we are already able
to approximate $\hat \Pi(\hat k)$ by a suitably chosen function.
Due to the
  non--local character of the integral equation, however, we have to know
the function $\hat \Pi(\hat k)$
for all momenta even if we were only interested in the
low momentum behavior of the self--energy. Consequently, each kind of
series ansatz is not a clever choice since it can never exhibit both,
the  correct small momentum behavior and the large $\hat k$ limit. We thus
employ a Pad\'e  (rational function)  approximation which will
turn out to be a remarkably good global approximation for $\Pi$.
The simplest ansatz consistent with $\hat \Pi'(0) =0$ and the asymptotic
behavior reads
\begin{equation} \hat \Pi (\hat k) = { A^2 + {\pi^2 \over 2 }
\kappa \hat k \over
a^2 + {a^2 \over A^2}{\pi^2 \over 2 } \kappa \hat k + \hat k^2 },\quad
A=A(\kappa ) , \; \; \; a=a(\kappa ) \lab{ans}
\end{equation}
and $a,A $ are functions of $\kappa $ which we have to determine such
that the approximation fits as well as possible the true solution.

 From the form of the denominators in the integrand (\ref{7a}) it is evident,
that $\hat \Pi (\hat p)$ contributes most to the integral for small values of
$\hat p$ since
for large values of  $\hat p$, $\hat \Pi$ being a decreasing function is
neglectable
compared with $\hat p^2$. For this reason we choose the ansatz
such that (\ref{7a}) is fulfilled exactly for $\hat\Pi (0)$ and $\hat\Pi''(0)$.
Plugging (\ref{ans}) into the integral equation, one could
--- at least in principle ---
carry out the integration
%for $\hat \Pi(0)$ and $\hat \Pi''(0)$
which results in two complicated transcendental equations for $a,A$ which
have to be solved numerically anyway.
We, however, will not do so, but merely discuss the limits $\kappa \to
0$ and $\kappa \to \infty$. It turns out (see appendix) that
for  large values of $\kappa$, the coefficients exhibit a  power law
behavior, namely $A(\kappa ) \sim A_0 \, \kappa^{2/3}, \; a(\kappa )
\sim a_0 \,\kappa^{1/3} $ with the constants $A_0 = 4,8611 \ldots $ and
$a_0 = 4,1140\ldots$.
For small $\kappa $ we find $A(\kappa )\sim \sqrt{6 \pi} \kappa^{1/2}$
and $a(\kappa ) \sim  \sqrt{12} + \ldots $. Values between the small
$\kappa$ and large $\kappa$ region can be found by numerical integration,
see also the table given in the appendix.
In Fig.\ 2 we compare the approximation  (\ref{ans})
with the exact one found
by numerical integration of Eq. (\ref{7a}).

\mkfig{
\begin{figure}
%\vspace*{-1cm}
\hspace*{\fill}
\epsfbox[0 50 288 248]{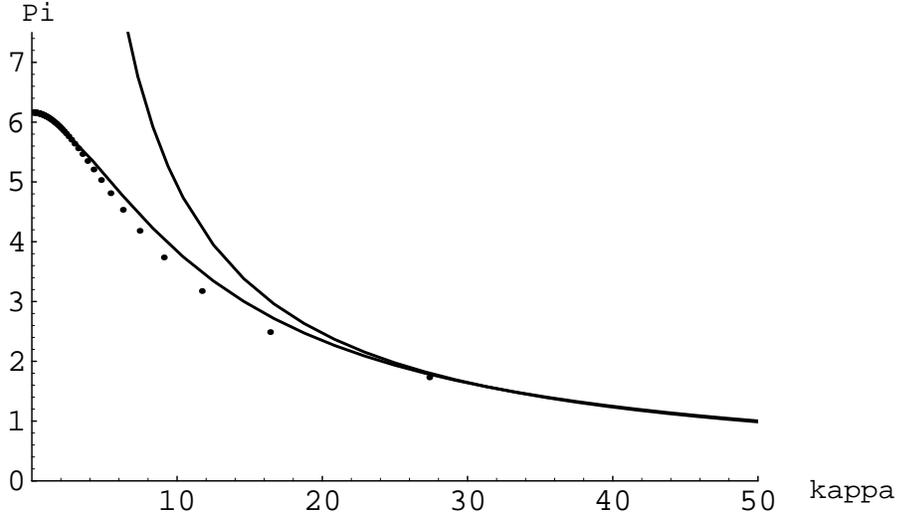}
\hspace*{\fill}
%\vspace*{-1cm}
\caption{Solution of the integral equation for $\hat \Pi$ at $\kappa=10$.
The dots represent values found by a $64$--point Gaussian quadrature of the
exact integral equation. The lower rigid line corresponds to the analytic
approximation of the self--energy employed throughout this paper.
We also indicate the exact asymptotic behavior of $\hat\Pi$ (upper line). }
\end{figure}
}

The solution for $\hat \Pi$ may further serve  to
eliminate the parameter $M$ as defined in (\ref{5}), which amounts to
evaluate the trace of the propagator,
\begin{equation}
T \sum_\omega \int { d^3 p \over (2 \pi )^3 } {1 \over p^2 + \omega^2 +
\mu^2 + \Pi( p,\omega ) } ,
\lab{8a} \end{equation}
where as before, we have to extract the UV--divergent part prior to
be able to make the static approximation $\omega =0$ in the sum.
Again subtracting and adding the UV--dominant contribution
\begin{equation} T \sum_\omega \int  { d^3 p \over (2 \pi )^3 } {1 \over p^2 +
\omega^2 + \mu^2} = {T^2 \over 12 } - {\mu T \over 4 \pi}
+ { \Lambda^2  \over 8 \pi^2 } + \mu^2 {\cal D }
,\lab{sub99} \end{equation}
%the difference
(\ref{8a}) -  (\ref{sub99}) is a
UV--finite quantity, and we are thus allowed to drop all modes except for
the zero--mode in the thermal sum. The difference
may be expressed by the function
\begin{equation}
\mu F(\kappa ) = {2 \over \pi}
\intl_0^\infty dp {p^2 \Pi (p) \over (p^2 + \mu^2 ) (
p^2 + \mu^2 + \Pi (p) ) } \lab{Fdef}
\end{equation}
which can be calculated analytically using the approximation for $\hat \Pi$
together with the known parameters $a$ and $A$.
We mention that, as before, also this function
exhibits a power law behavior  for
large values of the parameter $\kappa $, $ F(\kappa ) \sim F_0 \,
\kappa^{1/3} , \; F_0 = 0.9963 \ldots $, and goes like
 $F(\kappa ) \sim \kappa \pi /6$ for small $\kappa$.
Finally, putting together the definition of $M$ with the condition
(\ref{2a}), one can transform the  gap--equation
(\ref{5}) into an equation for  the mass parameter $\mu$,
\begin{equation}
\mu^2 + {\lambda^2 \phi^2 \over 2} {\cal D} = -m^2 + {\lambda \over 2}
\phi^2 + {\lambda \over 2} \left( {T^2 \over 12 }
- { \mu T \over 4 \pi }\left( 1  +  F(\kappa ) \right)+
{\Lambda^2 \over 8 \pi^2 } + \mu^2 {\cal D} \right) ,
 \lab{gap1}
\end{equation}
which is nevertheless meaningless until we absorb the divergent contributions
$\Lambda^2,\; {\cal D} $ in a redefinition of the bare quantities
$m,\lambda$ and $\phi$.

\section*{Renormalization }

The renormalization of $\lambda \Phi^4$--theory is delicate task
\cite{renp4}.
It can be shown that carrying out the limit in the regularization
parameter, the theory in fact becomes trivial in the sense that the
renormalized coupling constant vanishes.
For our purposes, however, it is more adequate to keep the
cut--off large but finite, since we want to study $\lambda\Phi^4$--theory
only as a model for the Higgs.
Consequently, we express the quantities $\lambda,\phi,m$ in redefined
ones,  but keep in mind
the eventual limit $\Lambda \to \infty$.

 We regroup the  corresponding terms (\ref{gap1})
in the following --- admittedly  suggestive --- way
$$ \mu^2 ( 1 -  {\lambda \over 2 }{\cal D } ) =
( -m^2 + \lambda {\Lambda^2 \over 16 \pi^2} ) + {\lambda \over 2} \phi^2
( 1- \lambda  {\cal D }) + \lambda \left( { T^2\over 24 } - {\mu T
\over 8 \pi } ( 1 + F(\kappa ) )  \right)
$$
and expect that in the renormalized equation, the coupling constant
$\lambda$  gets replaced by the renormalized one
$\lambda_R$ in the rightmost term.  Multiplying  the whole
equation by $\lambda_R / \lambda$ suggests the following
identifications
\begin{equation}
1 - {\lambda\over 2} {\cal D} = { \lambda \over \lambda_R} , \qquad
-m^2 + \lambda {\Lambda^2 \over 16 \pi^2} = { \lambda \over \lambda_R}
m_R^2  , \qquad
\phi^2 ( 1 - \lambda {\cal D} ) = \phi_R^2 .
\lab{rengl}
\end{equation}
Since we have only calculated to order $\lambda$ we are free to
neglect terms ${\cal O}(\lambda^2 ) $
which have to be fixed by higher order loop calculations.
Thus the equation for $\phi$ may be recast in the more convenient form
$$ \phi ( 1 - {\lambda\over 2} {\cal D} ) =  \phi_R
$$
which reveals an essential fact. From this equation and the
redefinition of the coupling constant $\lambda$ in (\ref{rengl})
we find immediately that
$\phi \lambda = \phi_R \lambda_R $ and consequently
$\kappa = \kappa_R$.
The invariance of $\kappa$ under the redefinitions is important for the
solutions of the gap--equation. As opposed to this, if $\kappa$ had a
non--trivial scaling with a divergent quantity, the integral
equation would have only the physically uninteresting solutions for
$\kappa=0$ or $\kappa \to \infty $ in the limit $\Lambda \to \infty$.
Finally we are able to reexpress the gap--equation for the mass
parameter in terms of the redefined quantities,
\begin{equation}
\mu^2 = -m_R^2 + {\lambda_R \over 2} \phi_R^2 + {\lambda_R \over 2}
\left( {T^2 \over 12 }  - {\mu T \over 4 \pi} (1 + F(\kappa ) )
\right) .
\lab{gap2}
\end{equation}

\vfill

\section*{Discussion}

\mkfig{
\begin{figure}
\epsfxsize5.7in
\vspace*{-1.5cm}
\hspace*{\fill}
\epsfbox{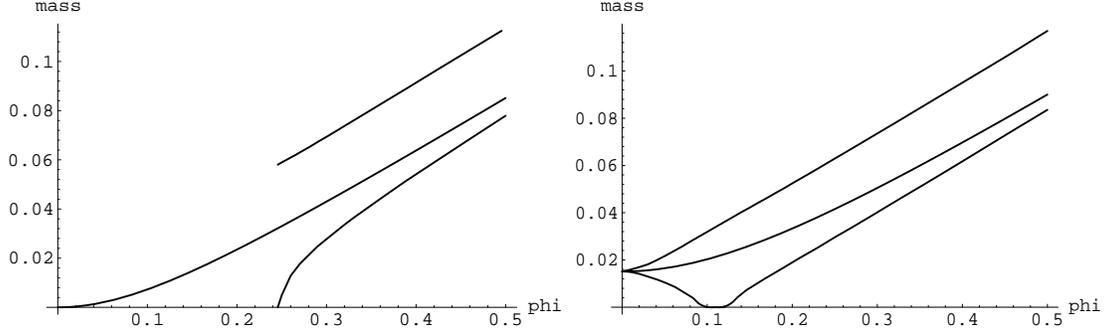}
\hspace*{\fill}
\vspace*{-1.2cm}
\caption{The mass parameter $\mu$ for the temperatures $T_c$ (left)
and  $\hat T$ (right) for $\lambda_R = 0.1 $ in units of $m_R$.
The lowest curve corresponds to the solution $\mu$ of the mass
equation. The middle curve are the values found without the correction
term $F(\kappa )$ (superdaisy resummation) and the upper curve
corresponds to the 'infrared mass'.  }
\end{figure}
}

The mass equation (\ref{gap2})
  --- together with the known function $F(\kappa )$ --- may be
used to determine the effective mass parameter $\mu$ for a given value
$\phi_R$. First we note that the  contribution $F(\kappa )$
entirely comes from the resummation of the
second graph in Fig.~1.
For small values of the parameter $\kappa$, the additional term
can be neglected with respect to $1$.
On the other hand, for large values of $\kappa$ which in the non--trivial
vacuum $\phi_R \neq 0 $ means small $\mu$,
due to the large $\kappa$ behavior of $F(\kappa ) \sim \kappa^{1/3}$,
the mass parameter cancels out in the next to leading
order contribution in equation (\ref{gap2}) and we are left with a term
 $\mu T F (\kappa ) \propto T^{4/ 3} (\lambda_R \phi_R )^{2/ 3}  $.
We emphasize that this term is an important contribution near the
critical temperature $T_c= m_R \sqrt{24 / \lambda_R} $ because
it takes on a non--vanishing value for $\mu \to 0$.
In particular, expressed in orders of $\lambda_R$,  since
$\phi_{R,min} = {\cal O}( 1 ) $ one finds $\lambda_R \mu T_c F(\kappa)
\sim \lambda_R
{T_c}^{4 / 3}
(\lambda_R \phi_R )^{2 / 3}={\cal O} ( \lambda_R )$ which is
of the same order as the term $\lambda_R \phi^2_R / 2 $ and thus
crucial at the critical temperature where the leading term
$-m_R^2 + \lambda_R T_c^2/24 $ in (\ref{gap2})
vanish.

Due to this behavior the mass--equation does not necessarily
have  a real solution for all values $\phi_R$.
In particular, at $T_c $ no solutions exists
for $ \phi_R \le m_R \sqrt{3} F_0^{3/4} / ( 8 \pi^5 )^{1/4} =0.245\ldots $.
For temperatures
between $T_c$ and $\hat  T=T_c \, ( 1 -\lambda_R F_0^{3/2}
/\sqrt{ 24 \pi^5} )^{-1} $, the solution consists of two disconnected branches
which join for $T\ge \hat T$ to give a solution for all values of
$\phi_R$ (Fig.~3).

The mass parameter may be interpreted physically
as a kind of 'ultraviolet'
mass appearing in the propagator (\ref{ans1}) since
the self--energy vanishes for
large momenta. As opposed to this, in the infrared region  $k \to  0$ the
self--energy does have the non--vanishing limit $\mu^2 A^2 / a^2 $
which contributes  to the mass in the propagator.
The corresponding effective masses are shown in Fig.~3.

\section*{Conclusion}

We have solved the gap--equation consistently to lowest order for a
spontaneously broken theory.  It turns out, that the Hartree--Fock
approximation corresponding to superdaisy resummation fails to be
consistent.
 Instead, for small effective mass, we encountered an additional contribution
in the gap--equation which behaves like $\phi^{2/3} $ which cannot
 be found in a perturbative calculation.
This contribution of course alters the effective propagator and
consequently the corresponding  effective potential.
Since it is
exactly the region of small mass which is of crucial importance
for the behavior of a theory near the phase transition, we expect significant
changes in the transition of field configurations from the false to the
trivial vacuum and vice versa.
 We plan to
clarify this point in a forthcoming investigation.

%\section{Tables}

%\acknowledgements

%Thank you, thank you

\newpage

\appendix
\section*{}

%\subsection{Large $\kappa$ behavior}

The conditions imposed to determine $a,A$ are
that equation (\ref{7a}) is fulfilled exactly for $\hat \Pi( 0 ) $ and
$ \hat \Pi ''( 0 ) $.
{}From the first condition, we find, putting the ansatz (\ref{ans}) into
(\ref{7a})
\begin{equation}
{A^2 \over a^2 } = \kappa \intl_{-1}^{1}  dz \intl_0^{\infty}
d\hat p { \hat p^2 \over N^2 } ,\quad
N=\hat p^2 +1 +  { A^2 + {\pi^2 \over 2 }  \kappa \hat p \over
a^2 + {a^2 \over A^2}{\pi^2 \over 2 } \kappa \hat p + \hat p^2 }. \lab{ah1}
\end{equation}
The functions $a,A$ can be approximated consistently for large values
of $\kappa$ by a simple power law, $A(\kappa ) \sim A_0 \, \kappa^{2/3}, \;
a(\kappa ) \sim a_0 \, \kappa^{1/3}$.
To see this explicitly, we plug the assumed behavior into equation
(\ref{ah1}) and rescale the momentum by $\hat p=\kappa^{1/3} x$
which transforms the denominator $N$ into
$$ N = 1 + \kappa^{2/3} L , \quad L= \left( x^2 +
{ A_0^2 + x {\pi^2 \over 2 } \over
a_0^2 + x { \pi^2 a_0^2 \over 2 A_0^2 } + x^2 }
\right). $$
For large values of $\kappa$ we neglect  1 with respect to
the term containing $L$. Thus the integral becomes proportional to
$\kappa^{-1/3}$ which is consistent with the power--behavior
of $A^2/a^2$  and  (\ref{ah1})   takes the form
\begin{equation}
 {A_0^2 \over a_0^2 } =  2 \intl_0^\infty dx { x^2 \over L^2  } . \lab{ah2}
\end{equation}
Similarly, from the second condition for $\hat \Pi''(0)$, one finds
\begin{equation}
%\lefteqn{
  {A_0^2 \over a_0^4 } =
\intl_0^{\infty} dx \left( { x^2 ( 3 + 2 L_2 + L_3 ) \over 3 L^3 }
- { 2 x^4 ( 2 + L_2)^2 \over 3 L^4 } \right)  , \lab{ah3} \end{equation}
$$  L_2= - { 2 A_0^4 ( 4 A_0^2 + \pi^2 x ) \over
( 2 a_0^2 A_0^2 + a_0^2 \pi^2 x + 2 A_0^2 x^2 )^2 } ,\quad
L_3 = { 8 A_0^6 ( -2 a_0^2 A_0^2 + 6 A_0^2 x^2 + x^3 \pi^2 )^2
\over ( 2 a_0^2 A_0^2 + a_0^2 \pi^2 x + 2 A_0^2 x^2 )^3 } .
$$
The set of equations (\ref{ah2},\ref{ah3}) can be solved numerically
for the constants $A_0,a_0$, and we find
$a_0=4.1140123\ldots ,\; A_0 = 4.8610750 \ldots .$

The function $F(\kappa )$ defined in (\ref{Fdef}) may be calculated
 analogously  by the same reasoning as above to have the asymptotic form
$F(\kappa) \sim F_0\, \kappa^{1/3}$ where the constant $F_0 =
0.996322285\ldots$
is given by
$$ F_0={2 \over \pi} \int_0^\infty dx { L-x^2 \over L}
$$

%\noindent  {\it  Small $\kappa$ behavior: }

For vanishing parameter $\kappa$, by the bounds given in (\ref{bounds}),
we have $\Pi_{max}  \sim \kappa$ and consequently
$$\hat \Pi(\hat k)  = \kappa {\pi \over\hat k} \arctan  {\hat k \over 2 } $$
Reexpressed in the functions $a,A$ this means
$A(\kappa )  \sim \sqrt{6 \pi} \kappa^{1/2}  $,
$ a(\kappa )  \sim \sqrt{12} $
and  $F(\kappa ) \sim \pi \kappa /6 $.

Further values may be read off from the following table.
%\noindent
\begin{center}
\begin{tabular}{|l|rrrrrrrrr|}
\hline
%$\kappa$ & $10^{-2}$ & $10^{-1}$ & 1&10&$10^2$&$10^3$&$10^4$&$10^5$&$10^6$ \\
%a &  3.491&3.70923&4.99502&9.09845&19.2453&41.5816 &89.5659&192.924&415.615\\
%A&0.435843&1.42065&5.2105&22.3065&104.917&492.957&2291.39&10636.5&49370.\\
%F
%%&0.0107933&0.0996745&0.657457&2.58646&7.40274&18.0866&41.1809&90.9804&198.296
$\kappa$ & $10^{-2}$ & $10^{-1}$ & 1&10&$10^2$&$10^3$&$10^4$&$10^5$&$10^6$ \\
\hline
a &  3.491&3.709&4.995&9.098&19.24&41.58 &89.56&192.9&415.6\\
A&0.4358&1.421&5.210&22.31&104.9&493.0&2291&10636&49370\\
F&0.005397&0.04984&0.3287&1.293&3.701&9.043&20.59&45.49&99.15
\\
\hline
\end{tabular}
\end{center}
% \newpage


\begin{references}
\bibitem{kolbt} E.~W.~Kolb and M.~S.~Turner, {\it The Early Universe},
(Addison--Wesley, Reading, MA, 1990);
A.~G.~Cohen, D.~B.~Kaplan and A.~E.~Nelson, Annu.~Rev.~Nucl.~Part.~Sci.\
43 (1993) 27.
\bibitem{stev}
I.~Stancu and P.~M.~Stevenson,  {\it Phys.~Rev.\ } {\bf D42}
(1990) 2710; \\
N.~Banerjee and S.~Mallik, {\it Phys.~Rev.\ } {\bf D43} (1991) 3368.
\bibitem{latt} See for example: C.~M.~Bender and F.~Cooper,
{\it Nucl.~Phys.} {\bf B224} (1983) 403; M.~L\"uscher, {\it Nucl.~Phys.} {\bf
B318} (1989) 705.
\bibitem{kirzl}
D.~Kirzhnits, {\it Pris'ma Zh.~Eksp.~Teor.~Fiz.} {\bf 15} (1972)
374 [JETP Lett.\ {\bf 15} (1972) 529];
D.~Kirzhnits and A.~Linde, {\it Phys.~Lett.\ } {\bf 42B} (1972) 471.
\bibitem{dwk}
L.~Dolan and R.~Jackiw, {\it Phys.~Rev.\ } {\bf D9} (1974) 3320;
S.~Weinberg, {\it Phys.~Rev.\ } {\bf D9} (1974) 3357;
D.~Kirzhnits and A.~Linde, {\em Zh.~Eksp.~Teor.~Fiz.} {\bf 67} (1975)
1263 [Sov.~Phys.\ JETP {\bf 40} (1975) 628].
\bibitem{CJT}
J.~M.~Cornwall, R.~Jackiw and E.~Tomboulis, {\it Phys.~Rev.\ }
{\bf D10} (1974) 2428.
\bibitem{pi}
G.~Amelino--Camelia and So--Young Pi, {\it Phys.~Rev.\ } {\bf D47}
(1993) 2356, unfortunately this paper suffers from a number of misprints.
\bibitem{8} K.~Takahashi, {\em Z.~Phys.~C } 26 (1985) 601;\\
M.~E.~Carrington, {\em Phys.~Rev.\ } {\bf D45 } (1992) 2933.
\bibitem{9} P.~Arnold and O.~Espinosa, {\em Phys.~Rev.\ }
{\bf D47} (1993) 3546.
\bibitem{renp4}
C.~Arag\~ao de Carvalho, S.~Caracciolo and J.~Fr\"ohlich, Nucl.~Phys.\
{\bf B215} (1987) 209, and references therein.
\end{references}
\end{document}